# GPCALMA: a Grid-based tool for Mammographic Screening

S. Bagnasco[a)], U. Bottigli[b)], P. Cerello[a], S. C. Cheran[c)], P. Delogu[d)], M. E. Fantacci[d)], F. Fauci[e)], G. Forni[f)], A. Lauria[f)], E. Lopez Torres[g)], R. Magro[e)], G. L. Masala[b)], P. Oliva[b)], R. Palmiero[f)], L. Ramello[h)], G. Raso[e)], A. Retico[d)], M. Sitta[h)], S. Stumbo[b)], S. Tangaro[i)], E. Zanon[j)]

a) INFN, Sez. di Torino, Italy.
b) Struttura Dipartimentale di Matematica e Fisica, Universita' di Sassari, and INFN, Sez. di Cagliari, Italy.
c) Dipartimento di Informatica, Universita' di Torino and INFN, Sez. di Torino, Italy.
d) Dipartimento di Fisica, Universita' di Pisa, and INFN, Sez. di Pisa, Italy.
e) Dipartimento di Fisica e Tecnologie Relative, Universita' di Palermo and INFN, Sez. di Catania, Italy.
f) Dipartimento di Scienze Fisiche, Universita' di Napoli and INFN, Sez. di Napoli, Italy
g) CEADEN, Havana, Cuba.
h) Dipartimento di Scienze e Tecnologie Avanzate, Universita' del Piemonte Orientale, and INFN, Sez. di Torino, Italy.
i) Dipartimento di Fisica, Universita' di Bari, and INFN, Sez. di Cagliari, Italy.
j) Ospedale Evangelico Valdese, Torino, and INFN, Sez. di Torino, Italy.

Corresponding Author: Piergiorgio Cerello
Address: INFN, Sezione di Torino, via P. Giuria 1, I-10125 Torino, Italy.
Tel.: +390116707416, fax: +390116699579, e-mail: cerello@to.infn.it

**Summary**
The next generation of High Energy Physics (HEP) experiments requires a GRID approach to a distributed computing system and the associated data management: the key concept is the *Virtual Organisation* (VO), a group of distributed users with a common goal and the will to share their resources. A similar approach is being applied to a group of Hospitals which joined the GPCALMA project (Grid Platform for Computer Assisted Library for MAmmography), which will allow common screening programs for early diagnosis of breast and, in the future, lung cancer. HEP techniques come into play in writing the application code, which makes use of neural networks for the image analysis and proved to be useful in improving the radiologists' performances in the diagnosis. GRID technologies allow remote image analysis and interactive online diagnosis, with a potential for a relevant reduction of the delays presently associated to screening programs.
A prototype of the system, based on *AliEn* GRID Services [1], is already available, with a central Server running common services [2] and several clients connecting to it. Mammograms can be acquired in any location; the related information required to select and access them at any time is stored in a common service called *Data Catalogue*, which can be queried by any client. The result of a query can be used as input for analysis algorithms, which are executed on nodes that are in general remote to the user (but always local to the input images) thanks to the PROOF facility [3], a set of C++ classes that provide the functionality required to configure several distributed nodes in a way that allows parallel analysis of similar data samples. The selected approach avoids data transfers for all the images with a negative diagnosis (about 95% of the sample) and allows an almost real time diagnosis for the 5% of images with high cancer probability.

**Introduction**
A reduction of breast cancer mortality in asymptomatic women is possible in case of early diagnosis, which is available thanks to screening programs, a periodical mammographic examination performed in general for 49-69 years old women.

The GPCALMA Collaboration aims at the development of tools that would help in the early diagnosis of breast cancer: Computer Assisted Detection (CAD) would significantly improve the prospects for mammographic screening, by quickly providing reliable information to the radiologists [4]. A dedicated software to search for massive lesions and micro-calcification clusters was developed recently (1998-2001): its best results in the search for massive lesions (micro-calcification clusters) are 94% (92%) for sensitivity and 95% (92%) for specificity [5].

Meanwhile, in view of the huge distributed computing effort required by the CERN/LHC collaborations, several GRID projects were started, mostly within the High energy Physics community. However, thanks to the involvement of some participants in medical physics projects, it became soon clear that the application of GRID technologies to a distributed database of mammographic images would facilitate a large-scale screening program, providing transparent and real time access to the full data set.

The data collection in a mammographic screening program will intrinsically create a distributed database, involving several sites with different functionality: data collection sites and diagnostic sites, i.e. access points from where radiologists would be able to query/analyse the whole distributed database. The scale is pretty similar to that of LHC projects: taking Italy as an example, a full mammographic screening program would act on a target sample of about 6.8 million women, thus generating 3.4 millions of mammographic exams/year. With an average data size of 60MB/exam, the amount of raw data would be in the order of 200TB/year: a screening program on the European scale would be a data source comparable to one of the LHC experiments (1-2PB/year).

GPCALMA was proposed in 2001, with the purpose of developing a "*GRID application*", based on technologies similar to those adopted by the CERN/ALICE Collaboration [6].

In each hospital, digital images associated to an exam will be stored on local resources and registered to a common service, known as *Data Catalogue*, which will preserve the mapping of the physical file names - required to actually access the data - to their logical file names, set by the users according to a pre-defined scheme, which must be meaningful to identify their properties.

Data describing the mammograms, also known as metadata, will also be stored in the *Data Catalogue* and, provided the set of metadata contains the parameters required to classify data, could be used to define an input sample for any kind of epidemiology study. The algorithm for the image analysis is sent to the remote site where images are stored, rather than moving images to the radiologist's sites. A preliminary selection of cancer candidates can be quickly performed and only mammograms with cancer probabilities higher than a selected threshold would be transferred to the diagnostic sites and interactively analysed by one or more radiologists.

Presently, a working version of the GPCALMA application is already available for local analysis. In parallel, a cluster of several PCs was already configured and successfully tested for the remote analysis of a set of mammograms. A dedicated *AliEn* Server is configured [2], with several Clients already attached in different sites:

tests of interactive dynamical input selection were successfully performed and a prototype Graphic User Interface (*GUI*) to the *Data Catalogue* was developed.
Making use of the *AliEn-PROOF* C++ Application Program Interface, now available in a prototype version to developers, it will soon be possible to integrate the *GUI*s to the algorithms and to the *AliEn* Services.

**The GPCALMA CAD Station**
The hardware requirements for the GPCALMA CAD Station are very simple: a PC running Linux with a SCSI bus connected to a planar scanner and to a high resolution monitor. The station can process mammograms directly acquired by the scanner and/or images from file and allows human and/or automatic analysis of the digital mammogram. The software configuration for the use in local mode requires the installation of ROOT [3] and GPCALMA, which can be downloaded either in the form of source code from the respective CVS servers. The functionality is usually accessed through a Graphic User Interface, or, for developers, the ROOT interactive shell. The Graphic User Interface (Fig. 1) allows the acquisition of new data, as well as the analysis of existing ones. Three main menus drive the creation of (access to) datasets at the patient and the image level and the execution of CAD algorithms. The images are displayed according to the standard format required by radiologists: for each image, it is possible to insert or modify diagnosis and annotations, manually select the Regions of Interest (*ROI*) corresponding to the radiologists geometrical indication. An interactive procedure allows zooming, either continuously or on a selected region, windowing, gray levels and contrast selection, image inversion, luminosity tuning.
The human analysis produces a diagnosis of the breast lesions in terms of kind, localization on the image, average dimensions and, if present, histological type.
The automatic procedure finds the *ROI*'s on the image with a probability of containing an interesting area larger than a pre-selected threshold value.

**Grid Approach**
The amount of data generated by a national or european screening program is so large that it can't be managed by a single computing centre. In addition, data are generated according to an intrinsically distributed pattern: any hospital participating to the program will collect a fraction of the total dataset. Still, that amount will increase linearly with time and, if fully transferred over the network to diagnostic centres, would be large enough to saturate the available connections.
However, the availability of the whole database to a radiologist, regardless of the data distribution, would provide several advantages:
- the CAD algorithms could be trained on a much larger data sample, with an improvement on their performance, in terms of both sensitivity and specificity;
- the CAD algorithms could be used as real time selectors of images with high breast cancer probability (see Fig. 2): radiologists would be able to prioritise their work, with a remarkable reduction of the delay between the data acquisition and the human diagnosis;
- data associated to the images (i.e., metadata) and stored on the distributed system would be available to select the proper input for epidemiology studies or for the training of young radiologists.

These advantages would be granted by a GRID approach: the configuration of a *Virtual Organisatio*n, with common services (*Data* and *Metadata Catalogue*, *Job Scheduler*, *Information System*) and a number of distributed nodes providing computing and storage resources would allow the implementation of the screening, tele-training and epidemiology use cases. However, with respect to the model applied to High Energy Physics, there are some important differences: the network conditions do not allow the transfer of large amounts of data, the local nodes (hospitals) do not agree on the raw data transfer to other nodes as a standard and, most important, some of the use cases require interactivity.

According to these restrictions, our approach to the implementation of the GPCALMA Grid application was based on two basic tools: *AliEn* [1] for the management of common services, *PROOF* [3] for the interactive analysis of remote data without data transfer.

**Data Management**
The GPCALMA data model foresees several Data Collection Centres (Fig. 2), where mammograms are collected, locally stored and registered in the *Data Catalogue*. In order to make them available to a radiologist connecting from a Diagnostic Centre, it is mandatory to use a mechanism that identifies the data corresponding to the exam in a site-independent way: they must be selected by means of a set of requirements on the attached metadata and identified through a Logical Name which must be independent of their physical name on the local hard drive where they are stored. *AliEn* implements these features in its *Data Catalogue* Services, run by the Server: data are registered making use of a hierarchical namespace for their Logical Names and the system keeps track of their association to the actual name of the physical files. In addition, it is possible to attach metadata to each level of the hierarchical namespace. The *Data Catalogue* is browsable from the *AliEn* command line as well as from the Web portal; the C++ *Application Program Interface (API)* to ROOT is under development. Metadata associated to the images can be classified in several categories: patient and exam identification data, results of the CAD algorithm analysis, radiologist's diagnosis, histological diagnosis, etc. Some of these data will be directly stored in the *Data Catalogue*, but some of them may be stored in dedicated files and registered: the decision will be made taking into account the radiologists' requirements. A dedicated *AliEn* Server for GPCALMA has been configured [2], in collaboration with the *AliEn* development team. Fig. 3 shows a screenshot from the WEB Portal. In addition, a Graphic User Interface (Fig. 4) drives the execution of three basic functionalities related to the *Data Catalogue*:
- registration of a new patient. It is based on the generation of a unique identifier, which could be easily replaced by the {\it Social Security} identification code;
- registration of a new exam, associated to an existing patient;
- query of the *Data Catalogue* to retrieve all the physical file names of exams associated to a given patient and eventually analyse them.

The top level functionality (interactive analysis of an image defined by the result of a query to the *Data Catalogue* regardless of its physical location) is not yet fully integrated in the GUI, but it is already provided via Command Line Interface and is described in the following section.

**Remote Data Processing**
Tele-diagnosis and tele-training require interactivity in order to be fully exploited, while in the case of screening it would be possible - altough not optimal - to live without. The PROOF (*Parallel ROOt Facility*) system [3] provides the functionality required to run interactive parallel processes on a distributed cluster of computers. A dedicated cluster of several PCs was configured and the remote analysis of digitised mammograms without data transfer was recently run.

The basic idea is that, whenever a list of input Logical Names is selected, it generates a list of physical names (one per image, consisting of the node name corresponding to the *Storage Element* where it is located and the physical path on its file-system. The information is used to dynamically generate a C++ script driving the execution of the CAD algorithms that analyse the image, which is sent to the remote node. Its output is a list of positions and probabilities corresponding to the image regions identified as possibly pathological by the CAD algorithms. Based on that, it is possible to decide whether the image retrieval is required for immediate analysis or not.

**Present Status and Plans**

The project is developing according to the original schedule. The CAD algorithms were rewritten in C++, making use of ROOT, in order to be PROOF-compliant; moreover, the ROOT functionality allowed a significant improvement of the Graphic User Interface, which, thanks to the possibility to manipulate the image and the associated description data, is now considered satisfactory by the radiologists involved in the project. The GPCALMA application code is available via CVS server for download and installation; a script to be used for the node configuration was developed. The *AliEn* Server, which describes the *Virtual Organisation* and manages its services, is installed and configured; some *AliEn* Clients are in use. The remote analysis of mammograms was successfully accomplished making use of PROOF.

Presently, all the building blocks required for the implementation of the tele-diagnosis and screening use cases were deployed and integrated into a prototype system.

As soon as the implementation of the data selection from the ROOT shell through the *AliEn* C++ *API* will be available, GPCALMA nodes will be installed in the participating hospitals and connected to the *AliEn* Server, hosted by INFN. It is foreseen that the task be completed by the end of 2004.

**Prospects**

The GRID approach to mammographic screening is very promising: it would grant the availability of a distributed database of mammograms from any node (i.e., hospital) in a *Virtual Organisation*. It would also allow to split the data collection and data analysis functions: data could be collected without the presence of radiologists, who would take advantage of the CAD results in order to select the sub-sample of mammograms with highest cancer probability and prioritise their work.

In principle, only mammograms considered positive by the CAD algorithm would have to be moved over the network to a different site, where a radiologist could analyze them, almost in real time.

However, given the present values of specificity obtained from the CALMA CAD algorithms, the amount of positive candidates would not be negligible and would amount to about 8-10% of the mammograms. These numbers would lead to a ratio of *CAD* positives to true positives which is still large: 8-10 assuming 1% of the collected mammograms are true positives.

Therefore the algorithm performance must be improved and part of our ongoing work is focused on that item. Still, our choice of moving the algorithm rather than the images would reduce the data transfer (i.e., the network requirements) by about a factor 10.

**Acknowledgments**
The authors wish to thank the *AliEn* development team for their support and guidance in the installation and configuration of the GPCALMA server.

**Bibliography**
1. http://alien.cern.ch.
2. http://gpcalma.to.infn.it.
3. http://root.cern.ch.
4. Ciatto S., Rosselli Del Turco M. et al., Brit. J. Cancer (2003) 85, 1645-1649.
   Karssemajer N., Otten JD. Et al., Radiology (2003) apr(227) (1) 192-200.
5. Lancet 2000, 355, 1822-1823.
6. http://alice.cern.ch.

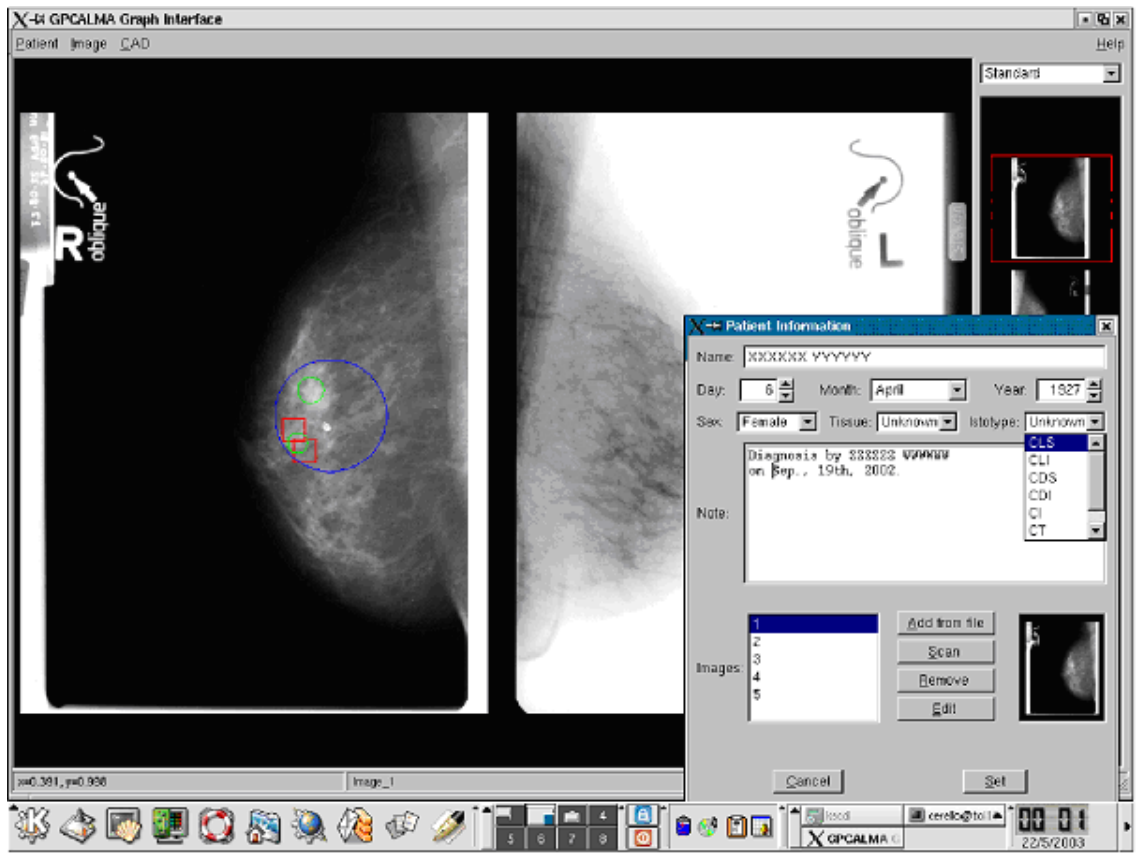

Fig. 1. The GPCALMA Graphic User Interface. Three menus, corresponding to the Patient, the Images and the CAD diagnosis levels, drive it. On the left, the CAD results for microcalcifications and masses are shown in red squares and green circles, together with the radiologist's diagnosis (blue circle). On the right, the image colours are inverted. The widget drives the update of patient and image related metadata.

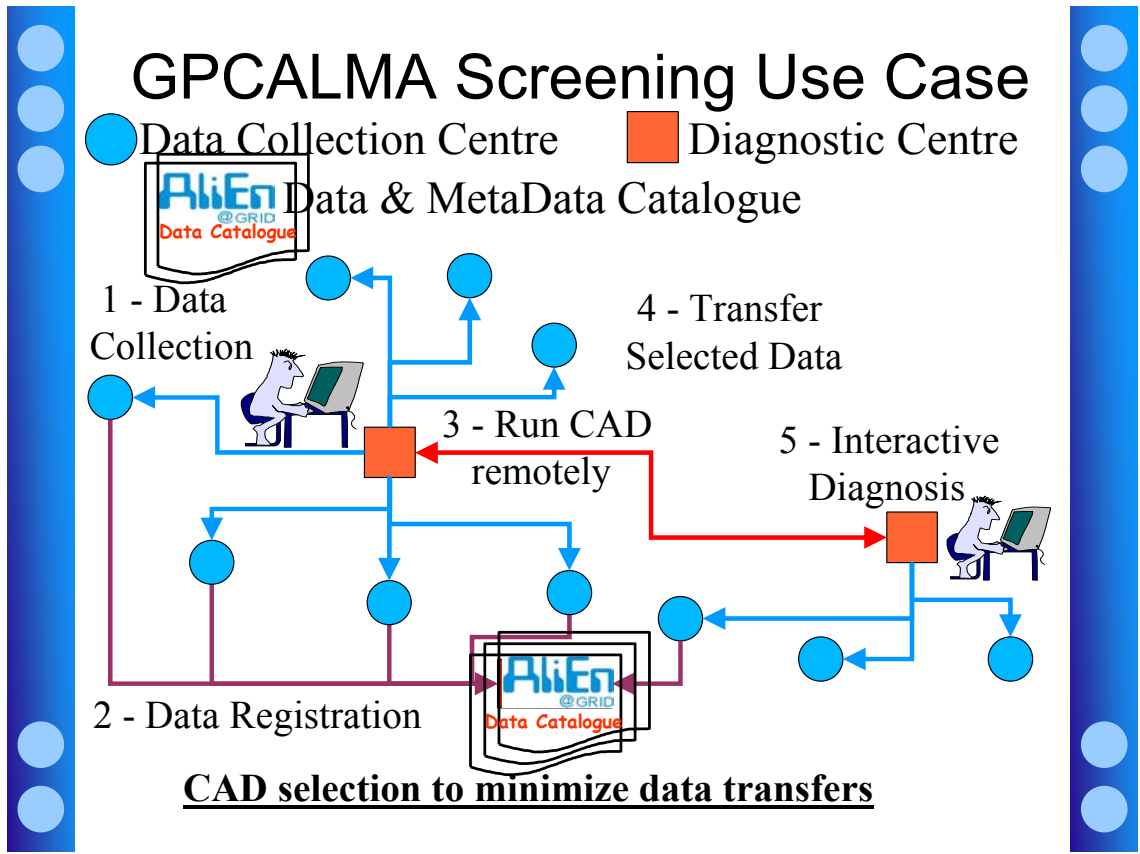

Fig. 2. The screening use case: Data Collection Centres store and register the images and the associated metadata in the *AliEn* Data Catalogue. Radiologists, from Diagnosis Centres, start the CAD remotely, without raw data transfer, making use of *PROOF*. Only the images corresponding to cancer probability larger than the selected threshold are moved to the Diagnosis Centre for the real-time visual inspection. Eventually, the small fraction of undefined cases can be sent to other radiologists.

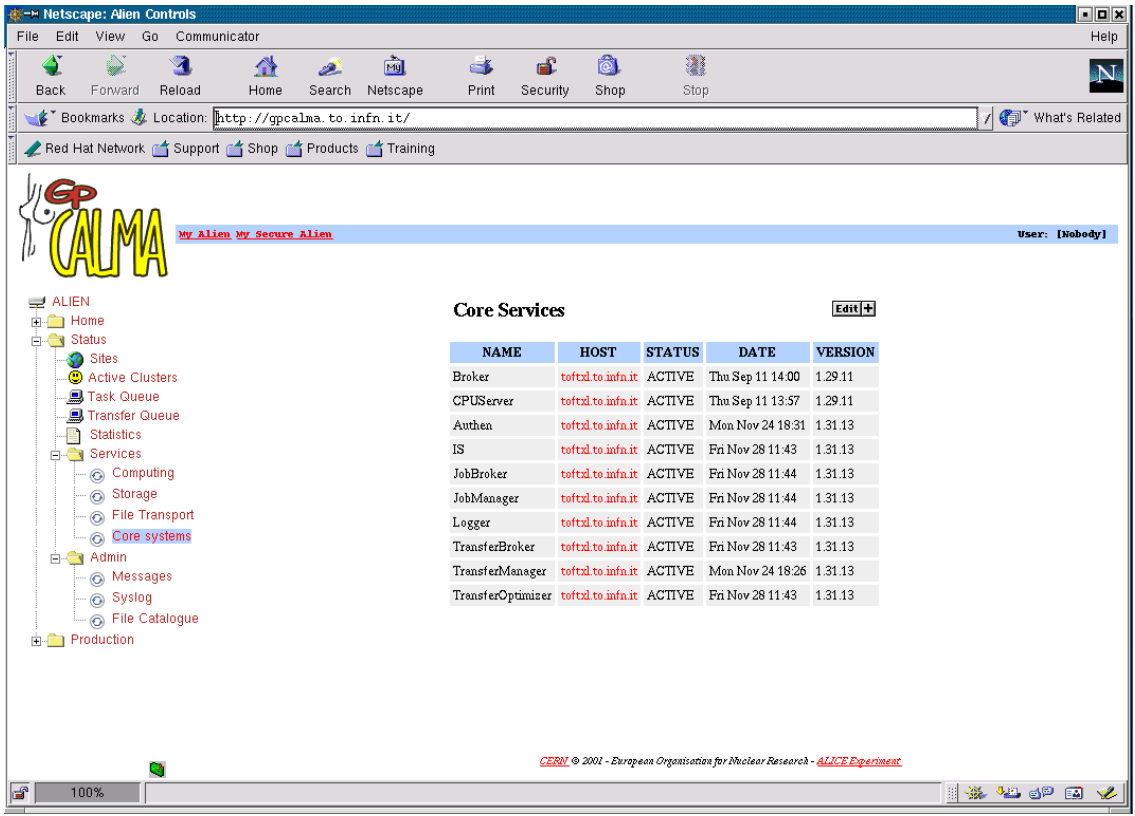

Fig. 3. Screenshot from the GPCALMA *AliEn* WEB Portal. Making use of the left side frame, the site can be navigated. General Information about the *AliEn* project, the installation and configuration guides, the status of the *Virtual Organisation* Services can be accessed. On the main frame, the list of the
core services is shown, together with their status.

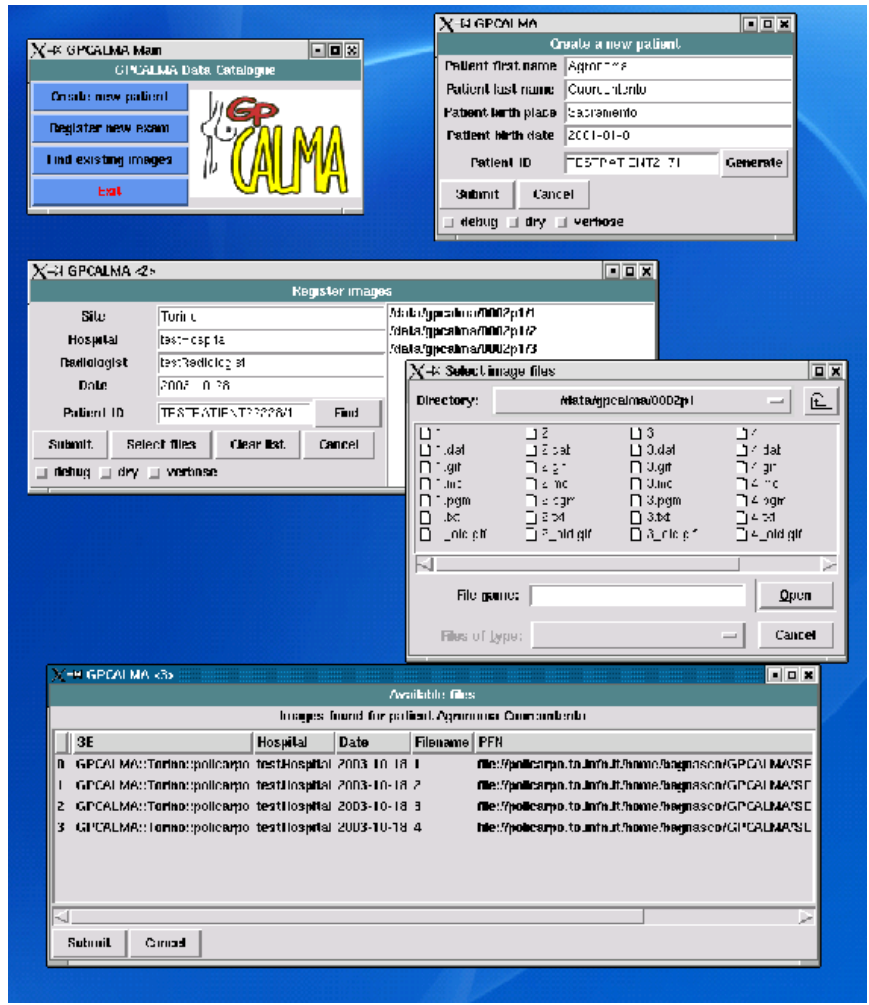

Fig. 4. Screenshot from the GPCALMA *AliEn* Graphic User Interface: the different widgets, driven by a main menu, show the basic functionality for the management of patient data and queries to the *Data Catalogue*: patient creation, association of new images (taken on the local filesystem) to an existing patient, retrieval of physical file names associated to images stored in the *Data Catalogue*, based on patient *MetaData*